\documentstyle[prl,twocolumn,aps,floats,epsf,epsfig,amsmath,amssymb]{revtex}
\begin{document}

\twocolumn[
\hsize\textwidth\columnwidth\hsize\csname@twocolumnfalse\endcsname
\draft

\title{Hall coefficient and magnetoresistance of 2D spin-polarized
  electron systems} 
\author{E. H. Hwang and S. Das Sarma}
\address{Condensed Matter Theory Center, 
Department of Physics, University of Maryland, College Park,
Maryland  20742-4111 } 
\date{\today}
\maketitle

\begin{abstract}

Recent measurements of the 2D Hall resistance show that the Hall
coefficient is independent of the applied in-plane magnetic field,
i.e., the spin-polarization of the system. We calculate
the weak-field Hall coefficient and the magnetoresistance of a spin
polarized 2D system using the semi-classical transport approach based
on the screening theory.  We solve the coupled 
kinetic equations of the two carrier system including
electron-electron interaction. We find that the in-plane magnetic
field dependence of the Hall coefficient is suppressed by the
weakening of  screening and the electron-electron interaction.
However, the in-plane magnetoresistance is mostly determined by the
change of the screening of the system, and can therefore be strongly
field dependent.


\end{abstract}
\vspace{0.5cm}
]

\newpage

The phenomena of apparent two dimensional (2D) metallic
behavior and the associated 2D metal-insulator transition (MIT) 
continue to attract a great deal of attention
\cite{d2,dsd_ssc}. 
The low temperature resistivity $\rho(T)$,
in zero applied field, shows remarkably strong ``metallic''-like
(i.e. $d\rho/dT >0$ for $n>n_c$) temperature dependence for 2D carrier
densities above the so-called critical carrier density ($n_c$) for the
2D MIT whereas, for $n<n_c$, the system exhibits insulating behavior
($d\rho/dT <0$). 
The application of an in-plane magnetic field $B$ 
has interesting effects on the 2D metallic phase, i.e., at a fixed low
$T$ the system develops a large positive magnetoresistance with $\rho(B)$
increasing very strongly (by as much as a factor of 4) with $B$ upto a
maximum field $B_s$, and for $B>B_s$, $\rho(B)$ either saturates (or
increases slowly with $B$ for $B>B_s$) showing a distinct kink at
$B=B_s$ \cite{si2}.
The observed temperature,
density, and parallel magnetic dependence 
of the 2D ``metallic'' resistivity \cite{dsd_ssc,dsh_b} can be
explained by the
screening theory in which the
strongly temperature dependent effective screened charged impurity 
disorder is
the qualitative reason underlying the striking metallic behavior
of dilute 2D carrier systems.

Remarkably,
recent measurements of the 2D Hall resistance \cite{vitkalov} 
in a parallel magnetic field have shown unexpected physical behavior
which is in sharp contrast with the strong in-plane field dependence
of the 2D magnetoresistivity.
The measured Hall coefficient seems to contradict qualitatively the results
based on the screening theory \cite{gold_b} even though the longitudinal
magnetoresistance can be explained by the change of the screening as
the spin-polarization of the system varies. The measured Hall coefficient is 
found not to vary with parallel magnetic field (or spin-polarization) for
fields ranging from 0 to well above $B_s$, where $B_s$ is the complete
spin-polarization field. However, the screening theory shows very
strong magnetic field (spin-polarization) dependence. Since the
screening theory, at least in its most elementary formulation
\cite{gold_b}, cannot explain this unexpected Hall coefficient data
Vitkalov {\it et al}. \cite{vitkalov} proclaim that the
electron-electron inter-subband scattering in the spin-polarized
system is the main reason for the experimental behavior.
The qualitative disagreement between the experimental Hall data of
ref. \onlinecite{vitkalov} and the screening theory \cite{gold_b} is
therefore a 
problem in our fundamental understanding of 2D transport because the
screening theory \cite{dsh} {\it can explain the temperature dependent Hall
coefficient}
in p-GaAs \cite{gao} and Si-MOSFET \cite{pudalov_hall}.

Motivated by this puzzling  experimental observation \cite{vitkalov}, we
investigate in this paper, based on the screening
model, the  spin-polarization dependence of the weak-field Hall
resistance and magnetoresistance.
For complete comparison with the experimental Hall coefficient we
include in our calculation the electron-electron scattering between
two different spin subbands. In ref. \onlinecite{vitkalov} Vitkalov
{\it et al}. compare the  Hall coefficient data with the
zero temperature results of screening theory \cite{gold_b} in the
strong screening 
limits $q_{TF}/2k_F \gg 1$ (where $q_{TF}$ is the Thomas-Fermi
screening wave
vector and $k_F$ is the Fermi wave vector). 
Our calculation, which includes finite temperature and
fully wave vector dependent
screening, is in qualitative agreement with
the experimental results \cite{vitkalov}
on the spin-polarization dependence of the weak
field Hall coefficient although electron-electron scattering might
play a role at finite temperature as we show. We therefore resolve the
experimental problem posed in ref. \onlinecite{vitkalov}.

To calculate the Hall coefficient and the magnetoresistance, we solve
the coupled kinetic equations for two kinds of carriers. 
When the parallel magnetic field is applied to the system the
electron densities $n_{\pm}$ for spin up/down are not equal with
the total density $n=n_+ + n_-$ fixed. 
The spin-polarized densities
themselves are obtained from the relative shifts 
in the spin up and down bands introduced by the Zeeman
splitting associated with the external applied field $B$. 
Since there are two groups of electrons (spin up and down) we need to consider 
inter-spin-band electron-electron scattering, which contributes to resistivity 
in addition to electron scattering by charged impurities and phonons.
In the presence of an applied field, the carrier momentum 
will relax to
equilibrium by electron-electron, electron-impurity, and
electron-phonon scattering 
(which is neglected in this calculation because at low
temperatures of interest to us phonon scattering is unimportant).
The electron-electron relaxation rate $1/\tau_{ee}$ will only affect
the relative momentum (i.e. the electron-electron scattering for the same
spin is neglected in this calculation). 
The electron-electron scattering relaxes the relative velocity or relative
momentum between two different populations to zero. 
In the steady state, 
the kinetic equations
of motion in the presence of an electric field {\bf E}
and magnetic field 
{\bf B} for spin up/down electrons, taking into account the collisions
with spin down/up electrons, have the form ($c=\hbar=1$) \cite{kukkonen}
\begin{eqnarray}
m_1\frac{{\bf v}_1}{\tau_1} + M\frac{n_2}{n}\frac{{\bf v}_1 -
  \bf{v}_2}{\tau_{ee}} & = &e {\bf E} + {e} ({\bf v}_1 \times {\bf
  B}), \nonumber \\ 
m_2\frac{{\bf v}_2}{\tau_2} + M\frac{n_1}{n}\frac{{\bf v}_2 -
  \bf{v}_1}{\tau_{ee}} & = &e {\bf E} + {e} ({\bf v}_2 \times {\bf
  B}), 
\end{eqnarray}
where $m_i$ is the effective mass ($i=1,2$ denotes up/down spin
subbands) for each group, $M =
nm_1m_2/(m_1n_1 + m_2n_2)$, and 
$\tau_i$ is the (energy and temperature dependent) 2D carrier
transport scattering time (the so-called momentum relaxation time)
determined by the {\it screened} charged impurity scattering
\cite{DH1} and $\tau_{ee}$ is
the electron-electron relaxation time
for the relative momentum of the spin polarized system. 

\begin{figure}
\epsfysize=2.3in
\centerline{\epsffile{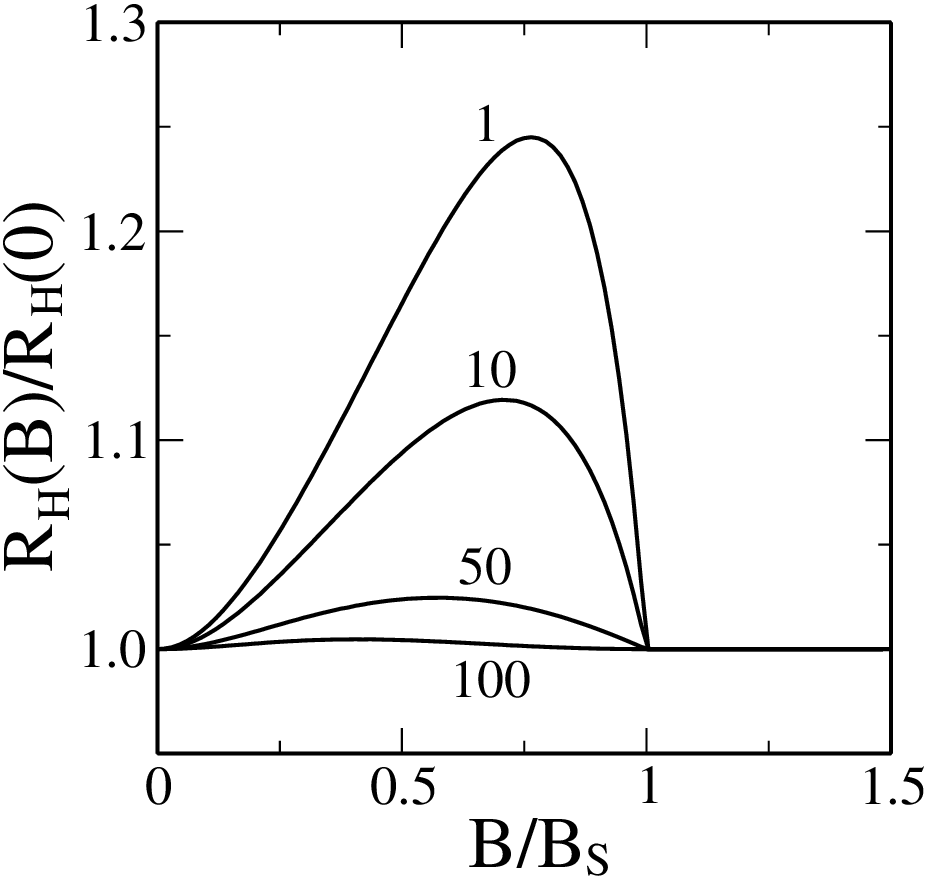}}

\caption{
Normalized Hall coefficient  $R_H(B)/R_H(0)$ as a function of in plane
magnetic field (spin polarization of the system) for different
densities, $n=1$, 10, 50, 100$\times 10^{10}cm^{-2}$ and at $T=0K$. 
At $B=B_s$ the 2D system is completely spin-polarized. 
}
\end{figure}

By solving the system of equation for ${\bf v}_i$ and substituting
these velocities into expression for the current density ${\bf j} =
n_+e{\bf v}_+ + n_2 e {\bf v}_-$, we find the resistivities
$\rho_{xx}$ and $\rho_{xy}$,
\begin{equation}
\rho_{xx}=\frac{1}{ne}\frac{\langle \tilde{\mu} \rangle 
 \left [ 1+ \frac{n_1\mu_2 +
      n_2 \mu_1}{n\mu_{ee}} 
  \right ]+\frac{\mu_1\mu_2(n_1\mu_2 + n_2 \mu_1)}{n}B_z^2}{ \left
    [ \langle \mu \rangle + \mu_1 \mu_2/\mu_{ee} \right 
  ]^2 + ( \mu_1   \mu_2 B_z)^2},
\end{equation}
\begin{equation}
\rho_{xy} = \frac{B_z}{ne} \langle r_H \rangle,
\end{equation}
where $\langle \tilde{\mu} \rangle = \langle \mu \rangle + \mu_1
\mu_2/\mu_{ee}$ and $B_z$ is the applied magnetic field normal to the
2D layer and $r_H$, the so-called Hall ratio, is given by
\begin{equation}
\langle r_H \rangle = 1 + \frac{ \langle \mu^2 \rangle - \langle \mu
  \rangle^2}
{ \left [ \langle \mu \rangle + \mu_1 \mu_2/\mu_{ee} \right ]^2 + ( \mu_1
  \mu_2 B_z)^2}.
\end{equation}
Here $\mu_i = e \tau_i/m_i$ and $\mu_{ee} = e\tau_{ee}/M$ and the average
mobility is defined by $ \langle \mu \rangle = (n_1 \mu_1 + n_2
  \mu_2)/{n} $ and $ \langle \mu^2 \rangle = (n_1 \mu_1^2 + n_2
  \mu_2^2)/{n}$.
When inter-subband electron-electron scattering is weaker than
the impurity transport scattering times ($\tau_{ee} \gg \tau_i$) we have
$\rho_{xx} = 1/ne\langle \mu \rangle$ and $\langle r_H \rangle =
\langle \mu^2 \rangle/ \langle \mu \rangle^2$.
In the other limit, $\tau_{ee} \ll \tau_i$, we have 
$\rho_{xx} = (n_1/n\mu_1 + n_2/n\mu_2)/ne$ and $\langle r_H \rangle
\rightarrow 1$. 
These equations immediately imply that the Hall coefficient will have
very weak temperature dependence (i.e. $r_H \sim 1$) when the
inter-spin subband electron-electron scattering dominates over the
impurity scattering.

\begin{figure}
\epsfysize=2.3in
\centerline{\epsffile{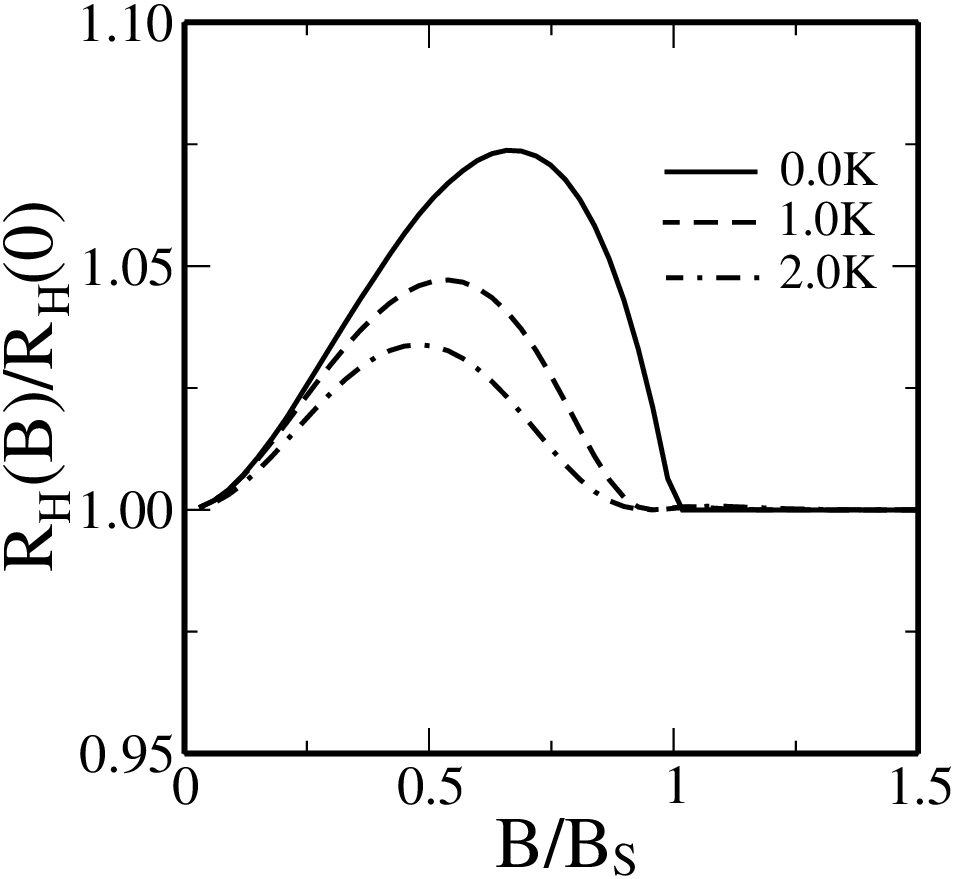}}

\caption{
Calculated Hall coefficient  $R_H(B)/R_H(0)$ as a function of in plane
magnetic field (spin polarization of the system) for different
temperatures, $T=0$, 1, 2K, and $n=20\times 10^{10}cm^{-2}$.
}
\end{figure}

In Figs. 1 and 2 we show our calculated Hall coefficient without
inter-subband electron-electron scattering. In this case the total
conductivity is 
the sum of the conductivities of each group, $\sigma = \sigma_1 +
\sigma_2$, and the Hall coefficient 
is given by $\langle r_H \rangle = \langle \mu^2 \rangle/\langle \mu
\rangle^2$.  Throughout this paper we use the parameters corresponding
to Si-MOSFET electron systems following ref. \onlinecite{vitkalov}. In
Fig. 1 we show our calculated Hall 
coefficient, $R_H(B)/R_H(0)$, ($R_H = \langle r_H \rangle/ne$) for
several carrier densities, $n=$1, 
10, 50, 100$\times 10^{10}cm^{-2}$ (which correspond to the
screening strength $q_{TF}/2k_F =$ 35, 11, 5, 3.5 respectively) as a
function of in-plane magnetic 
fields at $T=0$. As in-plane magnetic field increases the spins are
polarized, and at $B=B_s$ the system is completely spin-polarized.
In the low density limit (strong screening, $q_{TF}/2k_F \gg 1$) the
normalized Hall 
coefficient is strongly dependent on the polarization of the
system. But for high densities (weak screening) the coefficient is
almost independent of the spin polarization. 
Thus the polarization dependent Hall coefficient is suppressed as
screening effects decrease. We find that the normalized Hall coefficient
for a density $n=25 \times 10^{10}cm^{-2}$ (corresponding to
ref. \onlinecite{vitkalov}) 
increases only about 6\% at most as the system gets fully
spin-polarized by the applied field.

\begin{figure}
\epsfysize=2.3in
\centerline{\epsffile{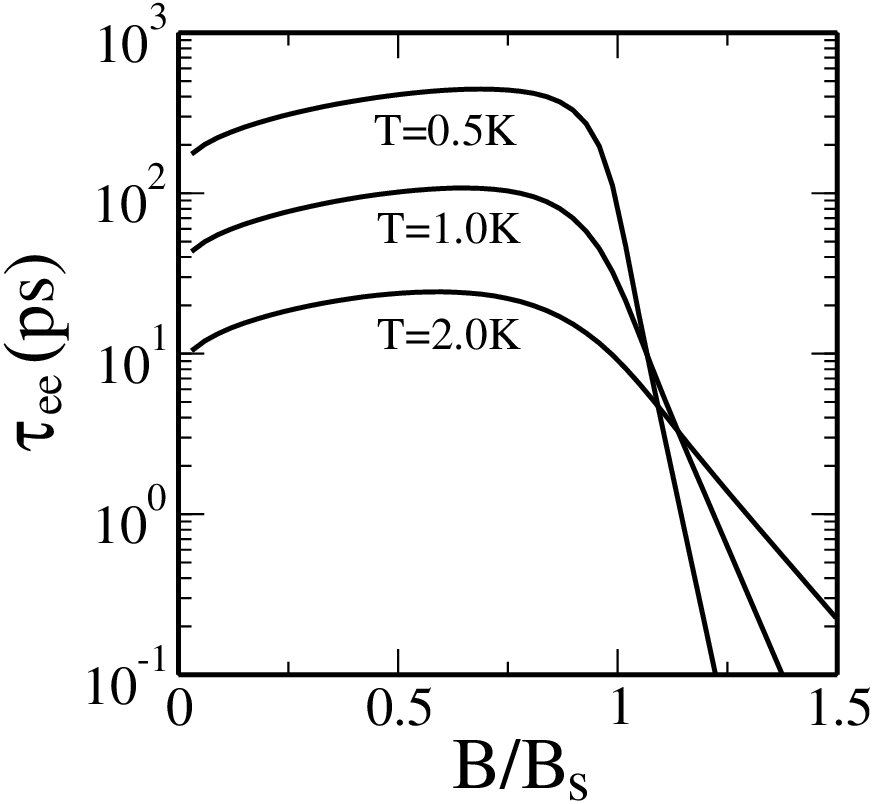}}
\caption{
Calculated inter spin-subband electron-electron scattering times for
different 
temperatures $T=$0.5, 1.0, 2.0K as a function of the parallel magnetic
field. We use the 2D density $n=20 \times 10^{10}cm^{-2}$.
}
\end{figure}

Since the screening function is suppressed by thermal effects 
we expect the Hall coefficient to be suppressed at
finite temperatures.
In Fig. 2 we show our calculated Hall coefficient, $R_H(B)$, 
for several temperatures, $T=$0, 1, 2K and a fixed density $n=20\times
10^{10}cm^{-2}$. As the temperature increases the normalized Hall
coefficient shows suppressed field dependence
mostly due to the weakening of the screening.
Comparison between our results 
and the experimental results \cite{vitkalov}
shows good qualitative agreement.
Thus, the observed field independence of
the Hall coefficient 
can be qualitatively explained by the screening theory. However, in
ref. \onlinecite{vitkalov} Vikalov {\it et al.} conclude, by comparing
their measured Hall coefficient (which does not vary with parallel magnetic
field) with the theoretical expectations based on the screening theory
\cite{gold_b} calculated in
the strong screening limit ($q_{TF}/k_F \gg 1$) and at zero
temperature, that the screening theory 
disagrees qualitatively with the experimental Hall effect results, and
therefore the strong
electron-electron scattering is a possible explanation for their data. 
In order to investigate the effects of electron-electron scattering on
the Hall coefficient we consider fully Eq. (4) which includes
electron-electron scattering as well as the screened charged impurities.
We explicitly calculate the electron-electron scattering time $\tau_{ee}$
defined as the relaxation time of the relative momentum between spin up and
down carriers \cite{appel}. For an unequal spin population we have $\tau_{ee}$ 
\begin{equation}
\frac{1}{\tau_{ee}} = \frac{8(k_BT)^2}{3 (\pi)^2}
\frac{nm^3}{n_1n_2}
p\int_{0}^{\pi}d\theta \sin\theta
\left | \frac{2\pi e^2}{q\varepsilon(q)}\right |^2 [f(p)]^2 
\end{equation}
where $f(p) = (1 + p^2 + 2p\cos\theta)^{1/2}$ with $p=(n_1/n_2)^{1/2}$,
$q=4\pi\sqrt{n_1n_2}\sin(\theta)/f(p)$, and
$\varepsilon(q)$ is the total dielectric function of the system.
Thus we have that $\tau_{ee}^{-1} \propto T^2$. In the very low
temperature limit we expect the contribution of the electron-electron
scattering to the Hall coefficient to be negligible because of this
$T^2$ dependence of $\tau_{ee}^{-1}$. At low temperature, therefore
($T \ll T_F$) our results (Figs. 1 and 2) neglecting $\tau_{ee}$
effects apply.

In Fig. 3 we show our calculated inter spin-subband electron-electron
scattering times, 
$\tau_{ee}$, as a function of the parallel field (or
spin-polarization). Our results show that $\tau_{ee}$ depends strongly
on the spin-polarization and temperature. (Note in
ref. \onlinecite{vitkalov} a constant parameter $\tau_{ee}$ is used
to fit the Hall coefficient data, which is incorrect.) In general, we find
the elastic scattering time due to ionized impurities at the interface
to be $\tau_i \approx 7$ps with an impurity density $n_i = 3 \times
10^{10}cm^{-2}$, which corresponds to the experimental Si-MOSFET sample of
the mobility $\mu \approx 2\times 10^4$ V/cm$^2$s.
Thus the calculated $\tau_{ee}$ is much larger than the elastic scattering time
$\tau_i$ in the low temperature limit where the 2D experiments are
typically carried out. 


\begin{figure}
\epsfysize=2.3in
\centerline{\epsffile{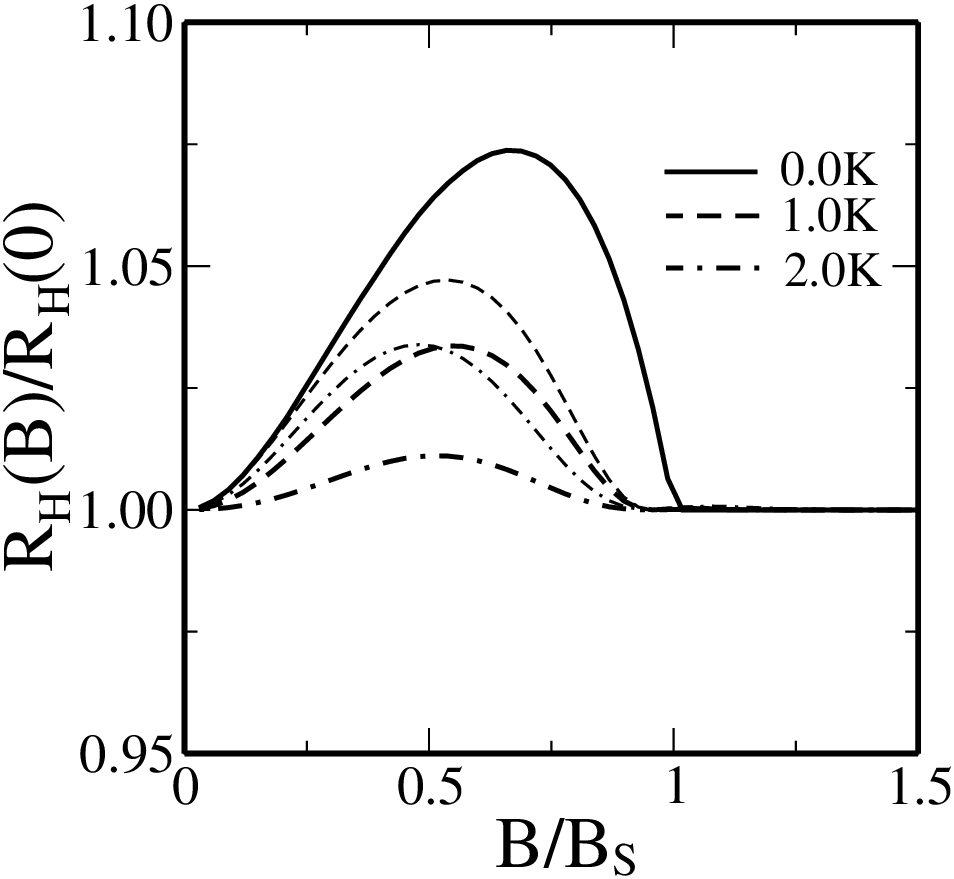}}
\caption{
Calculated Hall coefficient  $R_H(B)/R_H(0)$ as a function of in plane
magnetic field (spin polarization of the system) for different
temperatures, $T=0$, 1, 2K, and $n=20\times 10^{10}cm^{-2}$. Thick
(thin) lines indicate the results with (without) electron-electron
scattering. 
}
\end{figure}

With the results in Fig. 3 we calculate the Hall coefficient of the
spin-polarized system including electron-electron scattering. Fig. 4
shows the calculated Hall coefficient  $R_H(B)/R_H(0)$ as a function
of in plane magnetic field  for different
temperatures, $T=0$, 1, 2K, and $n=20\times 10^{10}cm^{-2}$. Our
calculation shows that the electron-electron scattering leads to the further
suppression in the field dependence of the Hall coefficient, 
leading to even better agreement between experiment \cite{vitkalov}
and our theory.


\begin{figure}
\epsfysize=2.3in
\centerline{\epsffile{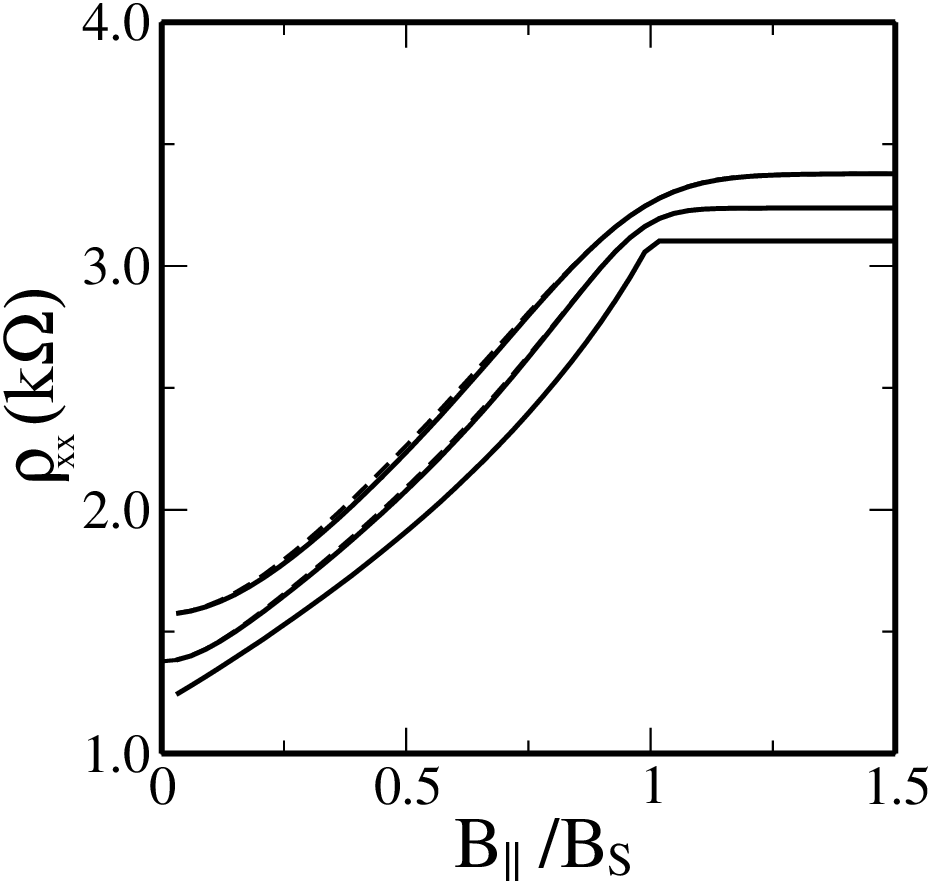}}
\caption{
Calculated magnetoresistance as a function of in-plane magnetic field
for different temperatures, T=0, 1, 2K (from bottom to top). The
results are shown for the carrier 
density $n=2 \times 10^{11}cm^{-2}$. The solid (dashed) lines indicate
calculated results without (with) electron-electron inter-spin subband
scattering.
}
\end{figure}


Fig. 5 shows the calculated magnetoresistance as a function of in-plane
magnetic field for different temperatures and a carrier density
$n=2\times 10^{11}cm^{-2}$. 
This strong positive magnetoresistance can be explained by  
the systematic suppression of screening as the spin-polarization
of the system changes \cite{dsh_b,gold_b}. The saturation of the
magnetoresistance above $B_s$ is attributed to the complete
spin-polarization of the system. Considering electron-electron
scattering in the calculation produces very small quantitative
modification in the
magnetoresistance even though the electron-electron scattering times
are comparable to the elastic impurity scattering times (at T=2K).
In Fig. 5 the solid (dashed) lines indicate
calculated results without (with) electron-electron interaction.
Even though the electron-electron scattering rate is stronger
than the elastic scattering rate, the calculated
magnetoresistance shows the same behavior. Thus, the experimentally
measured strong positive magnetoresistance with increasing  in-plane
magnetic field is induced mostly by the change of the screening
properties, not by the inter-spin subband electron-electron scattering.


\begin{figure}
\epsfysize=2.3in
\centerline{\epsffile{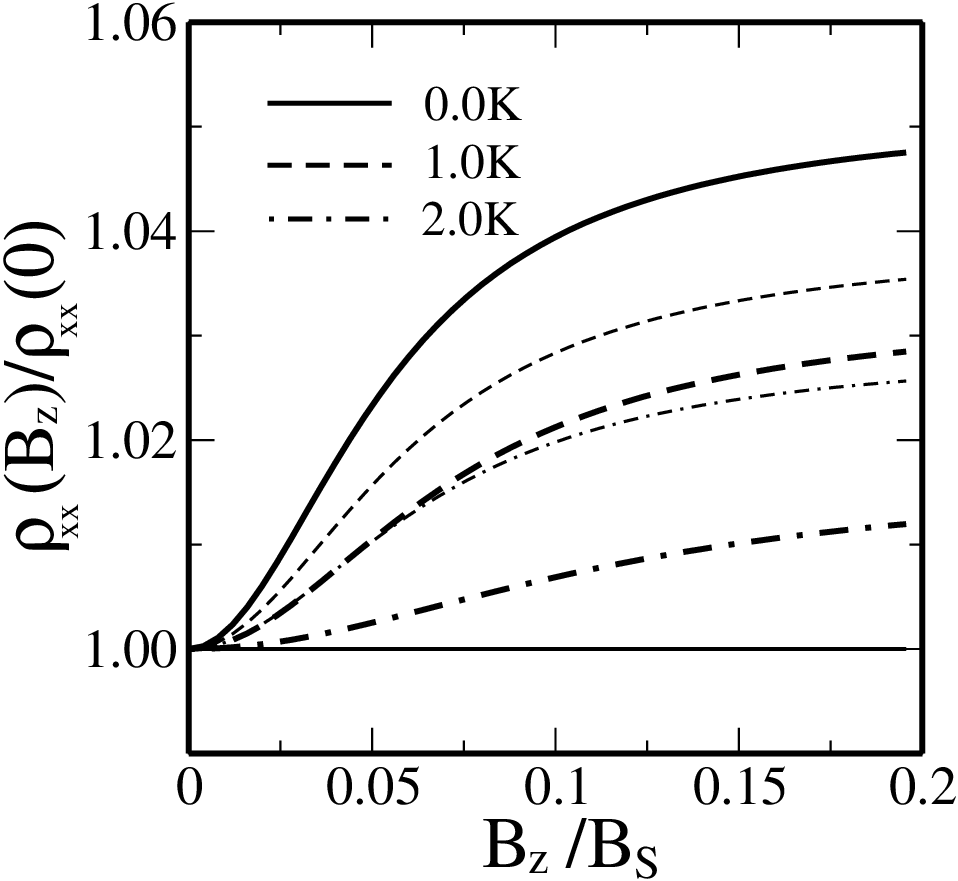}}
\caption{
Calculated magnetoresistance as a function of perpendicular magnetic
field ($B_z$) to the 2D plane for different temperatures, T=0, 1,
2K (from bottom). The results are shown for the carrier 
density $n=2 \times 10^{11}cm^{-2}$ and a fixed in-plane magnetic
field $B_{\|}=0.5B_s$. The thick (dashed) lines represent
results with (without) electron-electron interaction.
}
\end{figure}


In Fig. 6 we show that calculated longitudinal magnetoresistance
$\rho_{xx}$ as a function of perpendicular magnetic 
field ($B_z$) for different temperatures, T=0, 1,
2K. The in-plane parallel magnetic field solely gives rise to the spin
polarization of 
the system. The results are shown for a carrier 
density $n=2 \times 10^{11}cm^{-2}$ and a fixed in-plane magnetic
field $B_{\|}=0.5B_s$. For an unpolarized system (i.e., a single
subband with an isotropic scattering rate)
classical transport theory predicts no magnetoresistance (horizontal
solid line in Fig. 6) since the
Hall field exactly compensates the Lorentz force and the carriers
drift in the direction of the applied field. In fact, single subband
systems typically exhibit a negative magnetoresistance due to
quantum corrections arising from weak
localization or electron-electron interactions \cite{rmp}. 
For a system with two different Fermi wave vectors (two occupied
subbands) the classical transport theory predicts a positive
magnetoresistance varying quadratically with 
magnetic field $B_z$ at low fields and saturating at higher fields.
In Fig. 6 we have the positive magnetoresistance as the magnetic field
increases for a spin-polarized system. The positive magnetoresistance
is again reduced by  thermal effects (weakening of  screening),
and also by the 
electron-electron scattering (assuming that there are no quantum
corrections).

In conclusion, we calculate
the weak-field Hall coefficient and the magnetoresistance of a spin
polarized system based on the screening theory.  
We find that the spin-polarization dependence of the Hall
coefficient is strongly suppressed by the weakening of screening.
The electron-electron inter-spin subband scattering gives rise to an
additional suppression of the Hall coefficient, but 
in the low temperature experimental regime
the inter-spin subband electron-electron scattering 
is not quantitatively important. Our theory provides a very good
explanation for recent experiments \cite{vitkalov} on the magnetic
field dependence of the 2D Hall coefficient.


This work is supported by US-ONR, NSF, and LPS.



\begin{thebibliography}{99}


\bibitem{d2}E. Abrahams, S. V. Kravchenko, and M. P. Sarachik,
Rev. Mod. Phys. {\bf 73}, 251 (2001); S. V. Kravchenko and
M. P. Sarachik, Rep. Prog. Phys. {\bf 67}, 1 (2004)


\bibitem{dsd_ssc} S. Das Sarma and E. H. Hwang, Solid State Comm. {\bf
    135}, 579 (2005).

\bibitem{si2}T. Okamoto, K. Hosoya, S. Kawaji, and A. Yagi,
  Phys. Rev. Lett. {\bf 82}, 3875 (1999). 

\bibitem{dsh_b} S. Das Sarma and E. H. Hwang, Phys. Rev. B {\bf 72},
  205303 (2005). 

\bibitem{vitkalov}S. A. Vitkalov {\it et al}., \prb {\bf 63}, 193304 (2001);
S. A. Vitkalov, \prb {\bf 64}, 195336 (2001).

\bibitem{gold_b}V. T. Dolgopolov and A. Gold, JETP Lett. {\bf 71}, 27
  (2000).



\bibitem{dsh} S. Das Sarma and E. H. Hwang, \prl {\bf 95}, 016401
  (2005).

\bibitem{gao}X. P. A. Gao {\it et al.}, \prl {\bf 93}, 256402 (2004).

\bibitem{pudalov_hall} A. Yu. Kuntsevich {\it et al.},
Pis'ma Zh. Eksp. Teor. Fiz. {\bf 81}, 502 (2005); cond-mat/0504475 (2005).


\bibitem{DH1} S. Das Sarma and E. H. Hwang, \prl {\bf 83}, 164 (1999);
Phys. Rev. B {\bf 69}, 195305 (2004).



\bibitem{kukkonen} C. A. Kukkonen and P. F. Maldague, \prb {\bf 19},
  2394 (1979).

\bibitem{appel} J. Appel and A. W. Overhauser, \prb {\bf 18}, 758
  (1978).


\bibitem{rmp}P. A. Lee and T. V. Ramakrishnan, Rev. Mod. Phys. {\bf
    57}, 287 (1985); D. Belitz and T. R. Kirkpatrick,
Rev. Mod. Phys. {\bf 66}, 261-380 (1994).










\end{thebibliography}
\end{document}